\documentclass[10pt,technote,draftclsnofoot,letterpaper,onecolumn]{IEEEtran}
\usepackage{amssymb}
\usepackage{setspace}
\usepackage{amsfonts}
\usepackage{cite}
\usepackage[cmex10]{amsmath}
\usepackage{amsthm}
\usepackage{txfonts}
\usepackage{cases}
\usepackage{ifpdf}
\ifCLASSINFOpdf
  \usepackage[pdftex]{graphicx,hyperref,color}
  \DeclareGraphicsExtensions{.pdf,.jpeg,.png}
\else
  \usepackage[dvips]{graphicx,hyperref,color}
  \DeclareGraphicsExtensions{.eps}
\fi

\theoremstyle{plain}
\newtheorem{thm}{Theorem}

\newtheorem{cor}{Corollary}

\begin{document}
\title{On the Degree of Freedom for Multi-Source Multi-Destination Wireless Network with Multi-layer Relays}
\author{Feng~Liu,~ 
        Chung~Chan,~ 
        Ying~Jun~(Angela)~Zhang 
\thanks{The authors are with the Department of Information Engineering, The Chinese University of Hong Kong. E-mail: feng.steven.liu@gmail.com, chungc.mit@gmail.com, yjzhang@ie.cuhk.edu.hk.}
}

\maketitle

\begin{abstract}
Degree of freedom (DoF) region provides an approximation of capacity region in high signal-to-noise ratio (SNR) regime, while sum DoF gives the scaling factor. In this correspondence, we analyse the DoF region and sum DoF for unicast layered multi-hop relay wireless networks with arbitrary number of source/destination/relay nodes, arbitrary number of hops and arbitrary number of antennas at each node. The result is valid for quite a few message topologies. We reveal the limitation on capacity of multi-hop network due to the concatenation structure and show the similarity with capacitor network. From the analysis on bound gap and optimality condition, the ultimate capacity of multi-hop network is shown to be strictly inferior to that of single-hop network. Linear scaling law can be established when the number of hops is fixed. At cost of channel state information at transmitters (CSIT) for each component single-hop network, our achievable scheme avoids routing and simplifies scheduling.

\end{abstract}


\section{Introduction}

For wireless network, it is important to find/build the scaling law, which measures the scalability of communication capacity \cite{gupta2000capacity,gupta2003towards,xie2004network,kulkarni2004deterministic,xue2006scaling,ebrahimi2007throughput}. With the development of multi-antenna theory and technology, researchers have found that linear scaling law \footnote{Unlike Gupta-Kumar type, here scaling law represents the theoretical scalability without considering the practical constraints such as node density / position distribution / location area. We discuss this type of scaling law in this article.} can be obtained in only spatial domain \cite{bolcskei2006capacity}. The scaling factor in single-user multiple antenna systems is often called by pre-log factor, multiplexing gain, or degree of freedom (DoF), representing the number of independent streams/submessages which can be simultaneously transmitted on a time-frequency resource unit. For multi-user systems, DoF region is used to characterize the mutual limitation on all messages and the capacity of the whole network is scaled by the sum DoF - sum of all DoF variables. Different link topology, message topology, antenna configuration will result in different DoF regions. In fact, DoF region provides an approximation of capacity region in high signal-to-noise ratio (SNR) regime. Thus DoF region provides more details than the scaling law for understanding a network.

Many results on DoF of single-hop wireless network have been obtained. For the single-link or point to point channel, its DoF is well known to be at most the minimal number of antennas at transmitter or receiver \cite{foschini1998limits,telatar1999capacity}. For the multiuser multiple access channel (MAC) \cite{tse2004diversity} and broadcast channel (BC) \cite{yu2004sum,vishwanath2003duality} the DoF region has also been obtained. With these simple scenarios, centralized processing can be done at the common end of the links and the spatial resource can be fully exploited. Nevertheless, for a network with multiple transmitters and multiple receivers, only distributed processing is practical, which causes DoF loss. Assuming global channel state information at transmitter (CSIT), interference alignment (IA) based schemes can increase the scaling factor to some extent. The idea of IA is to allocate the signal spaces for all messages such that the interferences at each receiver can be aligned/overlapped into a common space and the efficiency of spatial resources is improved. Some researchers have studied the interference channel (IC) \cite{jafar2007degrees,cadambe2008interference} and X channel/network \cite{maddah2008communication,jafar2008degrees,cadambe2009interference}. A general $M\times N$ X network ($M$ transmitters and $N$ receivers) \cite{cadambe2009interference} with single antenna at each node has a sum DoF of $\frac{MN}{M+N-1}$, showing linear scaling factor $\frac{c}{(c+1)^2}\frac{n^2}{n-1}=O(n)$ with $n=M+N$ if $\frac{M}{N}=c$. Without CSIT, the DoF region will collapse except that some special structures are imposed on the channel.

Multi-hop wireless network is the generalization of single-hop network. There are two types of multi-hop: layered and non-layered. Each layer contains some nodes which transmit or receive simultaneously. If each path from source to destination has equal hops, we call the multi-hop network as layered, otherwise non-layered. The intermediate layers can act only as relay, i.e.: original messages are generated only at the source layer and wanted only at the destination layer, or might send/receive their own messages. According to the number of nodes at the source/destination layer, we have single-source single-destination network and multi-source multi-destination network. The latter is more general and complicated than the former. Moreover, if each message is destined to only one sink node, we call it as unicast, otherwise multicast or broadcast. Many works on multi-hop network focus on diversity, throughput, delay, scheduling and routing. However, there are few results on DoF region of multi-hop network. The sum DoF of parallel relay network with only one layer of relay is obtained in \cite{bolcskei2006capacity} and alternatively proven in \cite{cadambe2009interference}.

In this correspondence, we consider the DoF region of layered multi-source multi-destination unicast network with intermediate layers acting only as relay, assuming global CSIT is available for each layer. Half duplex decode-and-forward relay strategy is chosen. Our contribution includes the following aspects.
\begin{enumerate}
  \item For multi-hop networks with single antenna configuration, we provide an achievable DoF region following the routine of \cite{cadambe2009interference}. This DoF region applies to quite a lot message topologies. The lower and upper bounds on the sum DoF (approximate capacity) reveals the limitation of concatenation structure. The bound gap is analysed and optimality condition for zero gap is obtained. To satisfy the optimality condition, it is interesting to find that the number of nodes at the relay layers needs not be be infinity and there are many patterns about the distribution of node number at different layers. We also discuss the ultimate sum DoF when the number of nodes at each relay layer goes to infinity. Linear scaling law is established for some scenarios.
  \item We extend the above results to multi-hop networks with arbitrary antenna configuration. To the authors' best knowledge, these are the most general results on DoF for the discussed multi-hop network.
\end{enumerate}

\section{System model}\label{st:model}
\input{fig1.TpX}
The system model is demonstrated in Fig. \ref{fig:model}. Each circle denotes a wireless node. Between the source (S) and destination (D), there are $K$ layers of relay (R). Nodes only at source layer want to send some messages to some nodes only at destination layer. There is no direct link between source and destination so the relay is exploited. Half duplex is used at each layer of relay. For simplicity, we denote the source nodes as layer 0 and the destination nodes as layer $K+1$. We use $V_k^m$ to denote the $m$th node at layer $k$. Communication is only allowed between neighbouring layers $V_k$ and $V_{k+1}$ subsequently from layer 0 to layer $K+1$. It takes time period $T_k$ for transmission between layer $k$ and $k+1$, $\forall k=0,1,\dots,K$, where $T_k$ is length of symbol block. There are $\lvert V_k \rvert$ nodes for layer $k, k=0,1,\dots,K+1$. The link topology between two neighbour layers is with full connection. No link exists between non-neighbour layers, i.e., their channel coefficients are zero. The assumption on channel coefficients between neighbouring layers is same as that in \cite{cadambe2009interference}: non-zero, finite, and drawn according to a continuous distribution. The additive white Gaussian noise at each receiver is assumed with unit variance. Each node at layer $k$ knows all channel information between layers $k$ and $k+1$. Each transmitter has average power constraint $\rho$.

Each source node has at least one message to be sent to at least one of the destination nodes, as shown by the solid arrow with square end in Fig. \ref{fig:model}. Denote the active message between source $i$ and destination $j$ by $W^{[ji]}$. Assume all the active messages are mutually independent. For message $W^{[ji]}$, its DoF $d_{ji}$ indicates the number of independent parallel streams/submessages. The DoF region is defined by the set of achievable degrees of freedom: 
\begin{eqnarray}
d_{V_{K+1}V_{0}}:=\Bigg\{d_{ji}\in \mathbb{R}_{+}: \forall j\in V_{K+1}, \forall i\in V_{0} &|&
\forall w_{ji}\in \mathbb{R}_{+},\nonumber\\ &&
\sum_{j\in V_{K+1}}{\sum_{i\in V_{0}}{w_{ji} d_{ji}}} \le
\lim \sup _{\rho\rightarrow \infty}{ \sup _{\left[\left(R_{ji}(\rho)\right)\right]\in \mathcal{C}(\rho)}
	\sum_{j\in V_{K+1}}{\sum_{i\in V_{0}}{\frac{w_{ji} R_{ji}(\rho)}{\log (\rho)}}}}
	\Bigg \}
\end{eqnarray}
where $R_{ji}(\rho)=\frac{\log(\lvert W^{[ji]}(\rho) \rvert)}{\kappa_0}$ is the rate of the codeword encoding the message $W^{{ji}}$, $\kappa_0$ is the length of the codeword, $\mathcal{C}(\rho)$ denotes the capacity region of the network. It should be pointed out that here $W^{[ji]}$ and $d_{ji}$ are only conceptual. Due to the lack of direct links between source and destination, these messages might not be transmitted in their original forms. To accomplish the transmission task, some processing might be needed to obtain the messages suited for transmission in the multi-hop relay network, shown by the dashed arrows in Fig. \ref{fig:model}.  

Since the scaling factor approximates the capacity of the network, without confusion we use capacity $C$ to refer to the maximum achievable sum DoF $\sum_{j\in V_{K+1},i\in V_0} {d_{ji}}$.

\section{Multi-hop network with single antenna configuration}\label{st:singleantenna}
In this section, we give analysis on the DoF for multi-hop network with single-antenna nodes. Decode-and-forward strategy is assumed at each relay layer.

\subsection{An Achievable DoF region}
The main result is shown by the following theorem.
\begin{thm}[Achievable DoF region]
	\label{thm:dof}
	$d_{V_{K+1} V_{0}}$ is achievable if
	\begin{subequations}\label{eq:dof}
		\begin{align}
			\sum_{ji}{d_{ji}} &\leq \alpha\\
			\sum_{j} {d_{ji}} &\leq \frac{\alpha}{\lvert{V_0}\rvert}  \qquad \forall i\in V_0 \label{eq:sourcedof}\\
			\sum_{i} {d_{ji}} &\leq \frac{\alpha}{\lvert{V_{K+1}}\rvert} \quad \forall j\in V_{K+1}\label{eq:destdof}
		\end{align}	
	\end{subequations}
	where
	\begin{align*}
	\alpha^{-1}  &:= \sum_{k=0}^{K} {\alpha_k^{-1}}\\
	\alpha_k &:= \frac{\lvert{V_k}\rvert\lvert{V_{k+1}}\rvert}{\lvert{V_k}\rvert+\lvert{V_{k+1}}\rvert-1}.
	\end{align*}~
\end{thm}

\begin{IEEEproof}
	Apply decode-and-forward scheme with symmetric rate allocation, equalities is achieved for \eqref{eq:dof} using result of single-hop X network. The detailed proof is given as follows.
	
	According to Theorem 2 \cite{cadambe2009interference}, the single-antenna single-hop X network with $M$ transmitters and $N$ receivers has sum DoF
	\begin{equation}
	d_{\Sigma}=\frac{MN}{M+N-1},
	\end{equation}
	which can be achieved perfectly or asymptotically by a symmetric and uniform DoF allocation
	\begin{equation}\label{eq:dofxc}
	d_{ji}=\frac{1}{M+N-1}, \forall j=1,2,\dots,N; \forall i=1,2,\dots,M.
	\end{equation}
	We use this result to show the achievability of the given DoF region. Note, the above result of single-hop X network depends on symbol extension model in the achievable scheme. However, this will not affect our results since we can put it into frequency domain.
	
	The achievable scheme operates in $K+1$ phases and each phase has a block length of $T_k, k=0,1,\dots, K$. 
	
	The first phase is for the message transmission between the source nodes and the first layer relay nodes. The key method is message splitting. Each active message is uniformly split into $\lvert{V_{1}}\rvert$ independent submessages $W^{[ji]}_{n}, \forall n\in{V_{1}}$. All submessages $W^{[ji]}_n, \forall j\in{V_{K+1}}$ from source $i$ to the $n$th node at the first-layer relay are merged into a new message $W^{[ni]}_{0}$. Such a message splitting and merging procedure is demonstrated by Fig. \ref{fig:demo}. Thus we obtain a single-hop $\lvert{V_{0}}\rvert\times\lvert{V_{1}}\rvert$ X network with single antenna at each node. The corresponding coding scheme is employed so $\frac{T_0}{\lvert{V_{0}}\rvert+\lvert{V_{1}}\rvert-1}\log(\rho)+o(\log(\rho))$ bits for each new message are transmitted. At each receiver, totally $\frac{T_0 \lvert{V_{0}}\rvert}{\lvert{V_{0}}\rvert+\lvert{V_{1}}\rvert-1}\log(\rho)+o(\log(\rho))$ bits are decoded and the messages $W^{[ni]}_{0}, \forall i\in{V_{0}}$ are correctly received. The achievability proof of the X network with single antenna requires equal DoF for each new message, which indicates all source nodes have equal outgoing sum DoF for the super messages $W^{[i]}_0, \forall i\in{V_{0}}$. In other words, the DoF resource should be uniformly allocated to each source node. Denote the achievable sum DoF of the whole network as $\alpha$ (will be determined later). We need
	\begin{equation}\label{eq:sourcecondition}
		\sum_{j\in V_{K+1}} {d_{ji}} \leq \frac{\alpha}{\lvert{V_0}\rvert},  \quad \forall i\in V_0.
	\end{equation}

	Now, we step into the second phase. Each node at the first relay layer merges the received messages $W^{[ni]}_{0}, i\in V_0$ as a super message $W^{[n]}_{1}, n\in V_1$ and then splits it uniformly into $\lvert{V_{2}}\rvert$ independent messages $W^{[mn]}_{1}, m\in V_2$ for each of the receivers. After that, the relay nodes just forward these messages to the next layer. Again we obtain a single-hop $\lvert{V_{1}}\rvert\times \lvert{V_{2}}\rvert$ X network with single antenna at each node. All the bits are organized into symbols with block length $T_1$. With \eqref{eq:dofxc}, the DoF is uniformly allocated for each message. So $\frac{T_1 }{\lvert{V_{1}}\rvert+\lvert{V_{2}}\rvert-1}\log(\rho)+o(\log(\rho))$ bits for each message $W^{[mn]}_1$ can be correctly transmitted. Because each relay node sends out all received bits and does not add new information, we have the following equation:
	\begin{equation}
	\frac{T_0 \lvert{V_{0}}\rvert}{\lvert{V_{0}}\rvert+\lvert{V_{1}}\rvert-1}=\frac{T_1 \lvert{V_{2}}\rvert}{\lvert{V_{1}}\rvert+\lvert{V_{2}}\rvert-1} \nonumber
	\end{equation}
	or equivalently
	\begin{equation}\label{eq:T2vsT1}
	\frac{T_1}{T_0}=\frac{\lvert{V_{0}}\rvert}{\lvert{V_{2}}\rvert}\frac{\lvert{V_{1}}\rvert+\lvert{V_{2}}\rvert-1}{\lvert{V_{0}}\rvert+\lvert{V_{1}}\rvert-1}.\nonumber
	\end{equation}
	Then these messages are correctly decoded at the corresponding receivers and then forwarded to next layer.
		
	Repeating the above argument to the following phases, all relay layers can correctly decode and forward all active messages until the destination nodes finally collect all their desired messages. Since all messages received are uniformly split from the early stages, it implies that the ingoing sum DoF for each of the destination nodes should be equal, i.e. the DoF resource is uniformly allocated among the destination nodes:
	\begin{equation}\label{eq:destinationcondition}
		\sum_{i\in V_{0}} {d_{ji}} \leq \frac{\alpha}{\lvert{V_{K+1}}\rvert} \quad \forall j\in V_{K+1}.
	\end{equation}
	
	Generally, at the $k$th phase ($1\le k\le K$), we have the following equation
	\begin{equation}\label{eq:TkvsTk1a}
	\frac{T_{k-1} \lvert{V_{k-1}}\rvert}{\lvert{V_{k-1}}\rvert+\lvert{V_{k}}\rvert-1}=\frac{T_{k}  \lvert{V_{k+1}}\rvert}{\lvert{V_{k}}\rvert+\lvert{V_{k+1}}\rvert-1}
	\end{equation}
	or equivalently
	\begin{equation}\label{eq:TkvsTk1}
	\frac{T_k}{T_{k-1}}=\frac{\lvert{V_{k-1}}\rvert}{\lvert{V_{k+1}}\rvert}\frac{\lvert{V_{k}}\rvert+\lvert{V_{k+1}}\rvert-1}{\lvert{V_{k-1}}\rvert+\lvert{V_{k}}\rvert-1}.
	\end{equation}
	
	There are total $\lvert{V_0}\rvert\lvert{V_1}\rvert$ new messages sent from the sources and $\frac{T_0 \lvert{V_0}\rvert\lvert{V_1}\rvert}{\lvert{V_0}\rvert+\lvert{V_1}\rvert-1}\log(\rho)+o(\log(\rho))$ bits are transmitted to the destination over $\sum_{k=0}^{K}{T_k}$ period. The whole network can achieve the following sum DoF
	\begin{eqnarray}\label{eq:alpha0}
	\alpha
	=\frac{\frac{T_0 \lvert{V_0}\rvert\lvert{V_1}\rvert}{\lvert{V_0}\rvert+\lvert{V_1}\rvert-1}}{\sum_{k=0}^{K}{T_k}}
	=\frac{\lvert{V_0}\rvert\lvert{V_1}\rvert}{\lvert{V_0}\rvert+\lvert{V_1}\rvert-1}\frac{1}{\sum_{k=0}^{K}{\frac{T_k}{T_0}}}.
	\end{eqnarray}
	From \eqref{eq:TkvsTk1}, we have for $1\le k \le K$
	\begin{eqnarray}\label{eq:TkvsT0}
	\frac{T_{k}}{T_{0}}&=&\frac{T_{1}}{T_{0}}\frac{T_{2}}{T_{2}}\dots\frac{T_{k}}{T_{k-1}}\nonumber\\
	&=&\frac{\lvert{V_{0}}\rvert}{\lvert{V_{2}}\rvert}\frac{\lvert{V_{1}}\rvert+\lvert{V_{2}}\rvert-1}{\lvert{V_{0}}\rvert+\lvert{V_{1}}\rvert-1} \frac{\lvert{V_{1}}\rvert}{\lvert{V_{3}}\rvert}\frac{\lvert{V_{2}}\rvert+\lvert{V_{3}}\rvert-1}{\lvert{V_{1}}\rvert+\lvert{V_{2}}\rvert-1}
	\dots
	\frac{\lvert{V_{k-1}}\rvert}{\lvert{V_{k+1}}\rvert}\frac{\lvert{V_{k}}\rvert+\lvert{V_{k+1}}\rvert-1}{\lvert{V_{k-1}}\rvert+\lvert{V_{k}}\rvert-1}
	\nonumber\\
	&=&\frac{\lvert{V_{0}}\rvert\lvert{V_{1}}\rvert}{\lvert{V_{k}}\rvert\lvert{V_{k+1}}\rvert}\frac{\lvert{V_{k}}\rvert+\lvert{V_{k+1}}\rvert-1}{\lvert{V_{0}}\rvert+\lvert{V_{1}}\rvert-1}
	\end{eqnarray}
	So
	\begin{eqnarray}\label{eq:sumTkvsT0}
	\sum_{k=0}^{K}{\frac{T_k}{T_0}}&=&1+\frac{T_1}{T_0}+\frac{T_2}{T_0}+\dots+\frac{T_{K}}{T_{0}}\nonumber\\
	&=&1+\frac{\lvert{V_{0}}\rvert\lvert{V_{1}}\rvert}{\lvert{V_{1}}\rvert\lvert{V_{2}}\rvert}\frac{\lvert{V_{1}}\rvert+\lvert{V_{2}}\rvert-1}{\lvert{V_{0}}\rvert+\lvert{V_{1}}\rvert-1}+\frac{\lvert{V_{0}}\rvert\lvert{V_{1}}\rvert}{\lvert{V_{2}}\rvert\lvert{V_{3}}\rvert}\frac{\lvert{V_{2}}\rvert+\lvert{V_{3}}\rvert-1}{\lvert{V_{0}}\rvert+\lvert{V_{1}}\rvert-1}
	+\dots
	+\frac{\lvert{V_{0}}\rvert\lvert{V_{1}}\rvert}{\lvert{V_{K}}\rvert\lvert{V_{K+1}}\rvert}\frac{\lvert{V_{K}}\rvert+\lvert{V_{K+1}}\rvert-1}{\lvert{V_{0}}\rvert+\lvert{V_{1}}\rvert-1}
	\nonumber\\
	&=&\frac{\lvert{V_{0}}\rvert\lvert{V_{1}}\rvert}{\lvert{V_{0}}\rvert+\lvert{V_{1}}\rvert-1}\left[\frac{\lvert{V_{0}}\rvert+\lvert{V_{1}}\rvert-1}{\lvert{V_{0}}\rvert\lvert{V_{1}}}
	+\frac{\lvert{V_{1}}\rvert+\lvert{V_{2}}\rvert-1}{\lvert{V_{1}}\rvert\lvert{V_{2}}}
	+\frac{\lvert{V_{2}}\rvert+\lvert{V_{3}}\rvert-1}{\lvert{V_{2}}\rvert\lvert{V_{3}}}+\dots
	+\frac{\lvert{V_{K}}\rvert+\lvert{V_{K+1}}\rvert-1}{\lvert{V_{K}}\rvert\lvert{V_{K+1}}}\right]
	\end{eqnarray}
	Combining \eqref{eq:alpha0}\eqref{eq:sumTkvsT0} we can simplify the expression for the sum DoF as
	\begin{eqnarray}\label{eq:sumDoF}
	\alpha&=&\frac{1}{\frac{\lvert{V_{0}}\rvert+\lvert{V_{1}}\rvert-1}{\lvert{V_{0}}\rvert\lvert{V_{1}}}
		+\frac{\lvert{V_{1}}\rvert+\lvert{V_{2}}\rvert-1}{\lvert{V_{1}}\rvert\lvert{V_{2}}}
		+\frac{\lvert{V_{2}}\rvert+\lvert{V_{3}}\rvert-1}{\lvert{V_{2}}\rvert\lvert{V_{3}}}+\dots
		+\frac{\lvert{V_{K}}\rvert+\lvert{V_{K+1}}\rvert-1}{\lvert{V_{K}}\rvert\lvert{V_{K+1}}}}.
		\end{eqnarray}
	Define $\alpha_k := \frac{\lvert{V_k}\rvert\lvert{V_{k+1}}\rvert}{\lvert{V_k}\rvert+\lvert{V_{k+1}}\rvert-1}$. The above equation is equivalent to 
	\begin{equation}
	\alpha^{-1}  = \sum_{k=0}^{K} {\alpha_k^{-1}}.
	\end{equation}
			
\end{IEEEproof}
 
Notify that $\alpha_k$ represents the maximum achievable sum DoF of the single-hop network between layers $k$ and $k+1$. Theorem \ref{thm:dof} shows that the achievable sum DoF of the whole network is equal to that of the concatenation connection of all the component single-hop networks, while each single-hop network can be regarded as a capacitor. It is obvious that the sum DoF of the whole network is less than that of each component single-hop network. Thus if $\sum_{k=0}^{K}{\alpha_k}$ is constant, $\alpha$ is maximized when $\alpha_k$ equals to each other. Furthermore, multi-hop is harmful to the scaling factor: the more hops, the less achievable sum DoF. So the number of hops should be decreased as many as possible. From the proof we know that the delay is increased as $T=\sum_{k=0}^{K}{T_k}$. Therefore, to send messages from source to destination via relay layers, multi-hop suffers from both capacity decrease and delay increase. This is the penalty for lack of direct links. Comparison between the communication network and the capacitor network will be interesting: the former decodes and forwards information/bits, while the latter collects and stores electricity. The results on DoF/capacity with concatenation structure for both networks imply that the information and electricity share similar physical nature. 

The source and destination DoF conditions \eqref{eq:sourcedof} \eqref{eq:destdof} for the achievability are natural and includes quite a lot of message topologies. It can be easily verified that the parallel relay network is a special case of our model. Table \ref{tb:model} shows some examples which satisfy these conditions for a network with $\lvert{V_0}\rvert=\lvert{V_{K+1}}\rvert=3$.
\begin{table}[!t]
	\caption{Examples of the source and destination DoF conditions.}
	\label{tb:model}
	\centering
	\begin{tabular}{c||l}
		\hline
		\bfseries {Index} & \bfseries {DoF relationship for all active messages}\\
		\hline\hline
		1 & $d_{11}=d_{22}=d_{33}$ \\\hline
		2 & $d_{11}=2d_{22}=2d_{23}=2d_{32}=2d_{33}$ \\\hline
		3 & $d_{11}=d_{33}=d_{21}=d_{12}=d_{32}=d_{23}$ \\\hline
		4 & $d_{11}=d_{22}=d_{33}=d_{21}=d_{12}=d_{31}=d_{13}=d_{32}=d_{23}$ \\\hline
	\end{tabular}
\end{table}
 
\subsection{Bound on capacity}
In this subsection, we provide the lower and upper bounds on capacity of the whole network.
\begin{thm}[Bounds on capacity]
	\label{thm:cb}
	The capacity $C$ is bounded as follows
	\begin{eqnarray}
	\alpha \leq C\leq \beta
	\label{eq:cb}
	\end{eqnarray}
	where
	\begin{align*}
	\beta^{-1}  &:= \sum_{k=0}^{K} {\beta_k^{-1}}\\
	\beta_k&:= \min\{\lvert{V_k}\rvert,\lvert{V_{k+1}}\rvert\}
	\end{align*}~	
\end{thm}

\begin{IEEEproof}
	The lower bound in \eqref{eq:cb} follows from Theorem~\ref{thm:dof}. 
	
	The upper bound is obtained by converting each relay layer $V_k$ to a super-node with $\lvert{V_k}\rvert$ antennas for $\forall k=1,2,\dots,K$. With single-relay node in each layer, decode-and-forward scheme is optimal because the destination cannot decode the message if the relay cannot. We briefly give the proof as follows.
	
	At the first phase $T_0$, we meet a multi-access channel whose maximum DoF is given by $\min\left(\lvert{V_0}\rvert,\lvert{V_1}\rvert\right)$. In the following phases $T_k, \forall k=1,2,\dots, K-1$, the channel becomes single link and the maximum DoF is $\min\left(\lvert{V_k}\rvert,\lvert{V_{k+1}}\rvert\right)$. Finally at phase $T_K$ it is a broadcast channel with maximum DoF $\min\left(\lvert{V_K}\rvert,\lvert{V_{K+1}}\rvert\right)$. Now the equation \eqref{eq:TkvsTk1a} becomes
	       	\begin{equation}\label{eq:TkvsTk1b}
	       	T_{k-1}\min\left(\lvert{V_{k-1}}\rvert,\lvert{V_{k}}\rvert\right)=T_{k}  \min\left(\lvert{V_k}\rvert,\lvert{V_{k+1}}\rvert\right), \forall k=1,2,\dots,K.
	       	\end{equation}
	Define $\beta_k:= \min\{\lvert{V_k}\rvert,\lvert{V_{k+1}}\rvert\}$. The above equation is equivalent to
	       	\begin{equation}\label{eq:TkvsTk1c}
	       	\frac{T_{k}}{T_{k-1}}=\frac{\beta_{k-1}}{\beta_{k}}, \forall k=1,2,\dots,K.
	       	\end{equation}
	The sum DoF is obtained by
	\begin{eqnarray}\label{eq:beta0}
	\beta
	=\frac{T_0 \beta_0}{\sum_{k=0}^{K}{T_k}}
	=\beta_0\frac{1}{\sum_{k=0}^{K}{\frac{T_k}{T_0}}}.
	\end{eqnarray}
	From \eqref{eq:TkvsTk1c}, we have for $1\le k \le K$
	\begin{eqnarray}\label{eq:TkvsT0b}
	\frac{T_{k}}{T_{0}}&=&\frac{T_{1}}{T_{0}}\frac{T_{2}}{T_{2}}\dots\frac{T_{k}}{T_{k-1}}\nonumber\\
	&=&\frac{\beta_{0}}{\beta_{1}} \frac{\beta_{1}}{\beta_{2}}
	\dots
	\frac{\beta_{k-1}}{\beta_{k}}
	\nonumber\\
	&=&\frac{\beta_{0}}{\beta_{k}}
	\end{eqnarray}
	So
	\begin{eqnarray}\label{eq:sumTkvsT0b}
	\sum_{k=0}^{K}{\frac{T_k}{T_0}}&=&1+\frac{T_1}{T_0}+\frac{T_2}{T_0}+\dots+\frac{T_{K}}{T_{0}}\nonumber\\
	&=&1+\frac{\beta_{0}}{\beta_{1}}+\frac{\beta_{0}}{\beta_{2}}
	+\dots
	+\frac{\beta_{0}}{\beta_{K}}
	\nonumber\\
	&=&\beta_0 \sum_{k=0}^{K}{\beta_k^{-1}}.
	\end{eqnarray}
	Bring it into \eqref{eq:beta0} we obtain the upper bound
	\begin{eqnarray}\label{eq:beta}
	\beta
	=\frac{1}{\sum_{k=0}^{K}{\beta_k^{-1}}}.
	\end{eqnarray}	

\end{IEEEproof}	

Remarks: 1) Same result can be obtained if we further view the source/destination layer as a single super node, which implies that the cooperation among source/destination nodes is not necessary to achieve the upper bound. 2) Some layers of relay need not to be viewed as a super node. For example, only converting all odd relay layers and the last relay layer (i.e., layers $k=1,3,5,\dots,K$) into super nodes respectively still gives the same result. 3) The upper bound still shows the limitation due to concatenation structure.

\subsection{Bound gap and optimality condition}
To further understand the multi-hop network, we discuss the bound gap, optimality condition, ultimate capacity, and fractional gap in this subsection.
\begin{thm}[Bound gap]
	\label{thm:cbgap}
	The inverse gap between the upper and low bounds can be bounded as follows
	\begin{eqnarray}
	\alpha^{-1}-\beta^{-1}=\sum_{k=0}^{K} {\frac{\min\left(\lvert{V_k}\rvert,\lvert{V_{k+1}}\rvert\right)-1}{\lvert{V_k}\rvert\lvert{V_{k+1}}\rvert}}
	\leq \sum_{k\in L} {\frac{1}{\max\left(\lvert{V_k}\rvert,\lvert{V_{k+1}}\rvert\right)}}
	\leq \frac{\lvert{L}\rvert}{\min_{k\in L} \lvert{V_k}\rvert}
	\label{eq:gap}
	\end{eqnarray}
where 
\begin{equation}
L:=\left\{k\in \{0,1,2,\dots,K\}:\min\left(\lvert{V_k}\rvert,\lvert{V_{k+1}}\rvert\right)>1\right\}. 
\end{equation}	
\end{thm}
\begin{IEEEproof}	
	Consider computing the gap between the bounds
	\begin{subequations}\label{eq:bgap}
	\begin{align*}
	\alpha^{-1} - \beta^{-1} 
	&= \sum_{k=0}^{K} {\left(\alpha^{-1}_k-\beta^{-1}_k\right)}\\
	&= \sum_{k=0}^{K} {\left(\frac{\lvert{V_k}\rvert+\lvert{V_{k+1}}\rvert-1}{\lvert{V_k}\rvert\lvert{V_{k+1}}\rvert}-\frac{1}{\min\left(\lvert{V_k}\rvert,\lvert{V_{k+1}}\rvert\right)} \right)}\\
	&= \sum_{k=0}^{K} {\frac{\min\left(\lvert{V_k}\rvert,\lvert{V_{k+1}}\rvert\right)-1}{\lvert{V_k}\rvert\lvert{V_{k+1}}\rvert}}\\
	&\leq \sum_{k\in L} {\frac{1}{\max\left(\lvert{V_k}\rvert,\lvert{V_{k+1}}\rvert\right)}}.
	\end{align*}
	\end{subequations}

    From the facts that
	\begin{align}\nonumber
	\frac{1}{\max_{k\in {L}}\left(\lvert{V_k}\rvert,\lvert{V_{k+1}}\rvert\right)} &\leq \frac{1}{\lvert{V_k}\rvert} \leq \frac{1}{\min_{k\in {L}}\lvert{V_k}\rvert}
	\end{align}
	we can obtain the upper bound
	\begin{eqnarray}\nonumber
	\alpha^{-1}-\beta^{-1} \leq \frac{\lvert{L}\rvert}{\min_{k\in L} \lvert{V_k}\rvert}.
	\end{eqnarray}	
\end{IEEEproof}

From Theorem \ref{thm:cbgap}, the absolute gap between the upper and lower bounds is
	\begin{eqnarray}
	\beta-\alpha=\alpha\beta\left(\alpha^{-1}-\beta^{-1}\right)
	\leq \alpha\beta\sum_{k\in L} {\frac{1}{\max\left(\lvert{V_k}\rvert,\lvert{V_{k+1}}\rvert\right)}}
	\leq \frac{\alpha\beta\lvert{L}\rvert}{\min_{k\in L} \lvert{V_k}\rvert}
	\label{eq:absgap}
	\end{eqnarray}

\begin{cor}[Optimality condition]
	\label{cor:opt}
	$C=\alpha=\beta$ iff $\forall k=0,1,\dots,K$ 
	\begin{equation}\label{eq:opt}
	\{\lvert{V_i}\rvert,\lvert{V_{i+1}}\rvert\} \cap \{1,\infty\} \neq \emptyset
	\end{equation}
	~
\end{cor}
\begin{IEEEproof}
	From \eqref{eq:gap}, $\alpha=\beta$ if and only if the following condition holds
	\begin{equation}
	\max\left(\lvert{V_k}\rvert,\lvert{V_{k+1}}\rvert\right) = \infty, \forall k\in L.
	\end{equation}
	With the definition of $L$, it is easy to verify the above condition is equivalent to \eqref{eq:opt}.
\end{IEEEproof}

To illustrate the above optimality condition, we give an example where $K=3$. Some instances of the optimality condition are listed in Table \ref{tb:example}.
\begin{table}[!t]
\caption{Instances of the optimality condition for $K=3$. `x' indicates arbitrary number of nodes}
\label{tb:example}
\centering
\begin{tabular}{c||c|c|c|c|c}
\hline
\bfseries {Index} & $\lvert{V_0}\rvert$ & $\lvert{V_1}\rvert$ & $\lvert{V_2}\rvert$ & $\lvert{V_3}\rvert$ & $\lvert{V_4}\rvert$\\
\hline\hline
1 & 1 & x & 1 & x & 1\\\hline
2 & 1 & x & $\infty$ & x & 1\\\hline
3 & 1 & x & $\infty$ & x & $\infty$\\\hline
4 & 1 & $\infty$ & x & $\infty$ & x\\\hline
5 & $\infty$ & x & $\infty$ & x & $\infty$\\\hline
6 & x & 1 & x & 1 & x\\\hline
7 & x & 1 & x & $\infty$ & x\\\hline
8 & x & $\infty$ & x & $\infty$ & x\\\hline
\end{tabular}
\end{table}

\begin{cor}[Ultimate capacity]
	\label{cor:uc}
	If $\lvert{V_k}\rvert=\infty$ for $k=1,2,\dots,K$, the network achieve the following ultimate capacity 
	\begin{eqnarray}
	C^{\mathrm{U}}=(\lvert{V_0}\rvert^{-1}+\lvert{V_{K+1}}\rvert^{-1})^{-1}
	\label{eq:cinf}
	\end{eqnarray}
	~
\end{cor}
\begin{IEEEproof}
By Corollary \ref{cor:opt}, it can be easily verified that $\lvert{V_k}\rvert=\infty$ for $k=1,2,\dots,K$ satisfy the optimality condition. Thus $C=\alpha=\beta$. Then from Theorem \ref{thm:cb}, if $\lvert{V_k}\rvert=\infty$ for $k=1,2,\dots,K$, we will have
	\begin{equation}
	\beta_k^{-1} = 0, \quad\forall k=1,2,\dots,K-1. 
	\end{equation}~	
Then 
	\begin{equation}
	\beta^{-1} =\beta_0^{-1}+\beta_K^{-1}
	= \lvert{V_0}\rvert^{-1}+\lvert{V_{K+1}}\rvert^{-1}
	\end{equation}~	
Notify that $\beta_k^{-1}\ge 0$. The upper bound $\beta$ is bounded by
	\begin{equation}
	\beta=\frac{1}{\sum_{k=0}^{K} {\beta_k^{-1}}}\le\frac{1}{\beta_0^{-1}+\beta_K^{-1}}.
	\end{equation}~	
Therefore, we achieve the ultimate capacity.
\end{IEEEproof}	

Remarks: 1) Even though the optimality condition can be satisfied when some layers of relay have finite number of nodes, e.g., the instances 1/6/7/8 in Table \ref{tb:example}, the ultimate capacity cannot be achieved for these cases. In fact, the condition in Corollary \ref{cor:uc} is not only sufficient but also necessary. 
2) In contrast with the single-hop $\lvert{V_0}\rvert \times\lvert{V_{K+1}}\rvert$ X network, the ultimate capacity of the above multi-hop network shows some performance loss:
\begin{equation}
\gamma = \frac{\frac{\lvert{V_0}\rvert\lvert{V_{K+1}}\rvert}{\lvert{V_0}\rvert+\lvert{V_{K+1}}\rvert}}{\frac{\lvert{V_0}\rvert\lvert{V_{K+1}}\rvert}{\lvert{V_0}\rvert+\lvert{V_{K+1}}\rvert-1}}=\frac{\lvert{V_0}\rvert+\lvert{V_{K+1}}\rvert-1}{\lvert{V_0}\rvert+\lvert{V_{K+1}}\rvert}=1-\frac{1}{\lvert{V_0}\rvert+\lvert{V_{K+1}}\rvert}
\end{equation}
which implies that there is $\frac{1}{\lvert{V_0}\rvert+\lvert{V_{K+1}}\rvert}$ percent DoF loss due to the lack of direct link between source and destination. This loss is quite large when the numbers of nodes at source and destination are small. The largest loss is 50\% when $\lvert{V_0}\rvert=\lvert{V_{K+1}}\rvert=1$, i.e. single source single destination multi-hop network. The more number of nodes at source/destination, the less DoF loss with multi-hop relays.
When $\lvert{V_0}\rvert$ or $\lvert{V_{K+1}}\rvert$ becomes large enough, the loss can be neglected.
3) The ultimate capacity implies that the multi-layer relay network with infinity nodes at each relay layer is equivalent to a two-hop network, or a concatenation connection of two channels: MAC and BC with a super relay node as the intermediate, where the super relay node has infinite number of antennas (in fact, $\max\left(\lvert{V_0}\rvert,\lvert{V_K+1}\rvert\right)$ antennas are enough). So the number of relay layers does not matter if there are infinite nodes at each relay layer.

\begin{cor}[Fractional gap]
	\label{thm:fgap}
	The fractional gap between the capacity upper and lower bounds is
	\begin{eqnarray}
	\frac{\beta-\alpha}{C} \leq \beta(\alpha^{-1}-\beta^{-1})\le\frac{\lvert{L}\rvert}{(\lvert{V_1}\rvert^{-1}+\lvert{V_{K+1}}\rvert^{-1})\min_{i\in L}\lvert{V_i}\rvert}
	\label{eq:fgap}
	\end{eqnarray}
\end{cor}
\begin{IEEEproof}
	The gap follows from
	\begin{align*}
	\frac{\beta-\alpha}{C}
	&\leq \frac{\beta-\alpha}{\alpha}\\
	&=\beta \alpha^{-1}-1\\
	&=\beta(\alpha^{-1}-\beta^{-1})\\
	&\le\frac{\lvert{L}\rvert}{(\lvert{V_1}\rvert^{-1}+\lvert{V_{K+1}}\rvert^{-1})\min_{i\in L}\lvert{V_i}\rvert}
	\end{align*}
	In the final step of the above expression we use Theorem \ref{thm:cbgap} and the fact that $\beta$ is upper bounded by the ultimate capacity.
\end{IEEEproof}

\subsection{Scaling law}
Based on the above results, we analyse the scaling laws for some scenarios. Denote the total number of nodes in the network as $n$. There are $K$ layers of relay. Denote $c_{k}=\frac{\lvert{V}\rvert_{k}}{\lvert{V}\rvert_{k-1}}, \forall k=1,\dots,K+1$ which indicates the ratio of numbers of nodes at neighbour layers. By computing the lower and upper bounds with different scenarios, we can obtain the corresponding scaling laws. The results are listed in Table \ref{tb:scalinglaw}. It is clear that linear scaling law can be established if $K$ and $c_k, \forall k=1,\dots,K+1$ are fixed, i.e., the numbers of nodes at all layers increase proportionally with the total number of nodes in the network. If some layers have constant number of nodes, we get a constant scaling factor, which is determined by the concatenation of corresponding single-hop networks. If each layer has fixed number of nodes, the scaling factor is proportional to the inverse of $n$. This implies the capacity will decrease to zero if the number of relay layers goes to infinity. Again, we see that multi-hop is harmful.

\begin{table}[!t]
	\caption{Scaling laws for some scenarios.}
	\label{tb:scalinglaw}
	\centering
	\begin{tabular}{c||l|l}
		\hline
		\bfseries {Index} & \bfseries {Scenario} &  \bfseries {Scaling law}\\
		\hline\hline
		1 & Fixed $K$, fixed ratios $c_k, \forall k=1,\dots,K+1$ & linear: $O(n)$ \\\hline
		2 & Fixed $K$, fixed $\lvert{V}\rvert_{k}, \exists k\in\{0,1,\dots,K,K+1\}$ & constant: $O(1)$ \\\hline
		3 & Fixed $\lvert{V}\rvert_{k}, \forall k=0,1,\dots,K,K+1$ & inverse: $O(1/n)$ \\\hline
	\end{tabular}
\end{table}

\section{Multi-hop Network with arbitrary antenna configuration}\label{st:multipleantenna}

In this section we generalize the results in Section \ref{st:singleantenna} to network with arbitrary antenna configuration. Denote the number of antennas for node $m, m=1,2,\dots,\lvert V_k\rvert$ at layer $k, k=0,2,\dots, K+1$ by $A_k^m$.

\subsection{An Achievable DoF region}
\begin{cor}[Achievable DoF region]
	\label{thm:dof2}
	$d_{V_{K+1},V_{0}}$ is achievable if
	\begin{subequations}\label{eq:dof2}
		\begin{alignat}{2}
			\sum_{ji}{d_{ji}} &\leq \alpha\\
			\sum_{j} {d_{ji}} &\leq \frac{\alpha A_0^i}{\sum_{m=1}^{\lvert{V_{0}}\rvert}{A_0^m}}  \quad& \forall i\in V_0\label{eq:sourcedof2}\\
			\sum_{i} {d_{ji}} &\leq \frac{\alpha A_{K+1}^j}{\sum_{m=1}^{\lvert{V_{K+1}}\rvert}{A_{K+1}^m}} \quad& \forall j\in V_{K+1}\label{eq:destdof2}
		\end{alignat}	
	\end{subequations}
	where
	\begin{align*}
	\alpha^{-1}  &:= \sum_{k=0}^{K} {\alpha_k^{-1}}\\
	\alpha_k &:= \frac{\sum_{m=1}^{\lvert{V_{k}}\rvert}{A_{k}^m}\sum_{m=1}^{\lvert{V_{k+1}}\rvert}{A_{k+1}^m}}{\sum_{m=1}^{\lvert{V_{k}}\rvert}{A_{k}^m}+\sum_{m=1}^{\lvert{V_{k+1}}\rvert}{A_{k+1}^m}-1}
	\end{align*}~
\end{cor}

\begin{IEEEproof}
	The trick used here is antenna splitting which regards each antenna as a virtual distributed node. Thus at phase $k,k=0,1,2,\dots,K$, we obtain a $\sum_{m=1}^{\lvert{V_{k}}\rvert}{A_{k}^m} \times \sum_{m=1}^{\lvert{V_{k+1}}\rvert}{A_{k+1}^m}$ single-hop X network. The achievable sum DoF can be obtained based on Theorem \ref{thm:dof}. The symmetric and uniform DoF allocation requires the sum DoF for each source/destination node should be proportional to its number of antennas. We omit the detail proof for brevity.
\end{IEEEproof}
	
Table \ref{tb:model2} shows some examples which satisfy these conditions for a network with $\lvert{V_0}\rvert=2$, $ \lvert{V_{K+1}}\rvert=3$, $A_{0}^1=2$ and $A_0^2=A_{K+1}^1=A_{K+1}^2=A_{K+1}^3=1$. 
\begin{table}[!t]
	\caption{Examples of the source and destination DoF conditions with multi-antenna node.}
	\label{tb:model2}
	\centering
	\begin{tabular}{c||l}
		\hline
		\bfseries {Index} & \bfseries {DoF relationship for all active messages}\\
		\hline\hline
		1 & $d_{11}=d_{21}=d_{32}$ \\\hline
		2 & $d_{11}=d_{32}=d_{21}-d_{31}=d_{21}-d_{12}$ \\\hline
		3 & $d_{11}=d_{21}=d_{31}=2d_{12}=2d_{22}=2d_{32}$ \\\hline
	\end{tabular}
\end{table}

It is interesting to point out that the source and destination DoF conditions \eqref{eq:sourcedof2} \eqref{eq:destdof2} for the achievability introduce some flexibility: by shutting down one or more antennas at some nodes, the achievable scheme can be applied to more message topologies. 
 
\input{fig2.TpX}
Example 1 in Table \ref{tb:model2} with $K=2$ and two single-antenna nodes at each relay layer is demonstrated in Fig. \ref{fig:demo}. At each layer, solid frame box denotes a physical node, while dashed box denotes a virtual distributed node. For each message box, its height represents the DoF for the corresponding message, while its width represents the block length. For the split message boxes, same marker style indicates the same receiver. The colour of the message box indicate the destination node. We notify that the last relay layer does not split but only reorganize the decoded messages. The ordering of the split messages for an original message should be carefully handled at each destination node. Nevertheless, scheduling is greatly simplified and routing is even not needed.

\subsection{Bound on capacity}
	\begin{thm}[Bounds on capacity]
		\label{thm:cb2}
		The capacity $C$ is bounded as follows
		\begin{eqnarray}
		\alpha \leq C\leq \beta
		\label{eq:cb2}
		\end{eqnarray}
		where
		\begin{align*}
		\beta^{-1}  &:= \sum_{k=0}^{K} {\beta_k^{-1}}\\
		\beta_k&:= \min\left(\sum_{m=1}^{\lvert{V_{k}}\rvert}{A_{k}^m},\sum_{m=1}^{\lvert{V_{k+1}}\rvert}{A_{k+1}^m}\right)
		\end{align*}~	
	\end{thm}
	
	\begin{IEEEproof}
		The lower bound in \eqref{eq:cb2} follows from Corollary \ref{thm:dof2}. 		
		The upper bound is obtained by converting each relay layer $V_i$ to a super-node with $\sum_{m=1}^{\lvert{V_{k}}\rvert}{A_{k}^m}$ antennas for $\forall i=1,2,\dots,K$. The following proof is same as that for Theorem \ref{thm:cb}. We omit the detail for brevity.
	\end{IEEEproof}	

\subsection{Bound gap and optimality condition}
We observe that the lower and upper bounds in Theorem \ref{thm:cb2} can be easily obtained by replacing $\lvert{V_{k}}\rvert$ with  $\sum_{m=1}^{\lvert{V_{k}}\rvert}{A_{k}^m}$ in the corresponding formula of Theorem \ref{thm:cb}. Then, the results about gap analysis and optimality condition for single-antenna node scenario can be directly applied here. 

\subsection{Scaling law}
From the above results, when the antenna configuration is determined, scaling laws are similar as that for networks the single antenna configuration. However, the antenna configuration itself provides additional dimension to obtain linear scaling law if we fix the number of nodes and hops in the network. For example, we can obtain $s$-fold of the scaling factor if the numbers of antennas at all nodes are concurrently increased $s$-fold. In fact, the lower bound on capacity is obtained by regarding each antenna as a virtual distributed node with single antenna, which equivalently increases the number of nodes in the network. 

\section{Conclusion}\label{st:conclusion}
We considered the DoF aspects for multi-source multi-destination wireless network with multi-hop relays. An achievable DoF region was provided. We revealed the limitation on the concatenation structure and show its similarity with the capacitor network. It was shown that multi-hop is harmful for capacity and delay. So the number of hops should be decreased as many as possible. By analysing the lower and upper bounds, we obtained the bound gap, ultimate capacity, and optimality condition for the achievable scheme. We showed the scaling factor of the network and built linear scaling law for some scenarios. At the cost of CSIT at each component single-hop network, routing is avoided and scheduling can be simplified.

\bibliographystyle{IEEEtran}
\bibliography{IEEEabrv,references}

\end{document}